\documentclass[]{elsart}
\usepackage{amssymb}
\usepackage{pst-node,pstricks,graphicx}
\usepackage{lineno}

\begin{document}                % INITIALIZE - DONT CHANGE

\flushright{v2.0}

\begin{frontmatter}
\title{
Geometry reconstruction of fluorescence detectors revisited
}

\author{D.~Kuempel,\corauthref{cor1}}
\author{K.-H.~Kampert}
\author{and}
\author{M.~Risse}
\corauth[cor1]{ {\it Correspondence to}:~D.~Kuempel
(kuempel@physik.uni-wuppertal.de)
}
\address{
Bergische Universit\"{a}t Wuppertal, Department of Physics, \\
Gau\ss str. 20, D - 42097 Wuppertal, Germany \\
}

%\linenumbers
%%%%%%%%%%%%%%%%%%%%%%%%%%%%%%%%%%%%%%%%%%%%%%%%%%%%%%%%%%%%%%%%%%%%%
\begin{abstract}                % DON'T CHANGE THIS LINE

The experimental technique of fluorescence light observation is used in
current and planned air shower experiments that aim at understanding
the origin of ultra-high energy cosmic rays.
In the fluorescence technique, the geometry of the shower is
reconstructed from the correlation between arrival time and
incident angle of the signals detected by the telescope.
The calculation of the expected light arrival time used so far in shower
reconstruction codes is based on several assumptions.
Particularly, it is assumed that fluorescence photons are produced
instantaneously during the passage of the shower front and that
the fluorescence photons propagate on a straight line with vacuum
speed of light towards the telescope.
We investigate the validity of these assumptions, how to correct them,
and the impact on reconstruction parameters when adopting realistic
conditions.
Depending on the relative orientation of the shower to the telescope,
corrections can reach 100~ns in expected light arrival time,
0.1$^\circ$ in arrival direction and 5~g~cm$^{-2}$ in depth of shower
maximum.
The findings are relevant also for the case of ``hybrid''
observations where the shower is registered simultaneously by
fluorescence and surface detectors.

\end{abstract}

\end{frontmatter}

%%%%%%%%%%%%%%%%%%%%%%%%%%%%%%%%%%%%%%%%%%%%%%%%%%%%%%%%%%%%%%%%%%%%%

%\linenumbers

\section{Introduction}
\label{sec-intro}

% MR: I will work on the introduction

Understanding the origin and nature of ultra-high energy (UHE) cosmic
rays above $10^{19}$~eV is a major challenge of astroparticle 
physics~\cite{reviews}.
These cosmic rays are studied by detecting the atmospheric showers
they initiate.
Current and planned air shower
experiments~\cite{auger,hires-gzk,telarray,euso,owl}
use the technique of
fluorescence light observation: shower particles deposit energy in
the atmosphere through ionisational energy loss. Part of this
energy (of order $10^{-4}$)
% \marginpar{Echt?!?})
is emitted isotropically at near-UV
wavelengths in de-excitation processes.
These fluorescence photons can be detected by appropriate
telescope systems operating in clear nights. 
Typically, pixel cameras with 25$-$100~ns timing resolution are
used, where an individual pixel covers a field of view of about
1$-$1.5$^\circ$ in diameter (see e.g.\ Ref.\ \cite{auger}).
The signal (light flux per time) is registered as a function of the
viewing direction of the pixels.

The first step to reconstruct the primary parameters of an observed
air shower is given by the determination of the shower geometry.
An accurate geometry reconstruction is, for instance, decisive for
directional source searches;
but it is also a prerequisite for reconstructing
other important shower parameters such as the primary energy or the
depth of shower maximum.
We note that also the shower energies obtained from Auger ground array
data are calibrated by the fluorescence telescopes~\cite{roth_icrc07}.

\begin{figure}[t]
\begin{center}
\includegraphics[width=10.0cm]{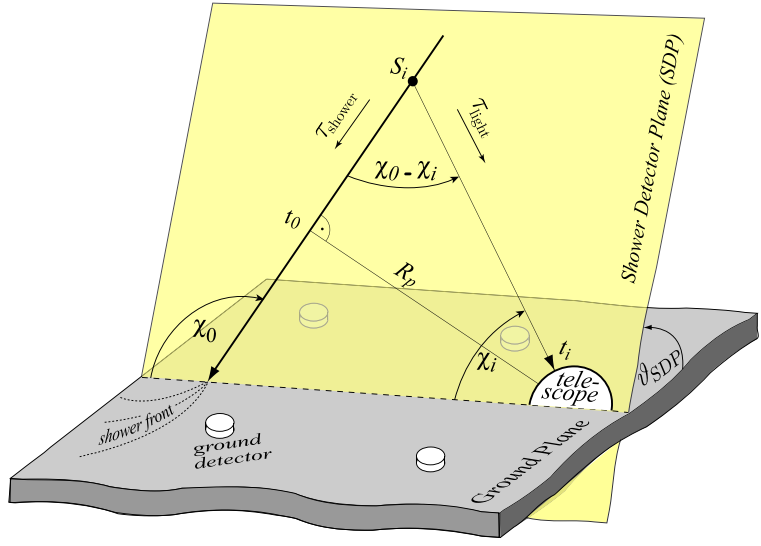}
\caption{
Sketch of the shower geometry and quantities used in the derivations.
}
\label{fig-sdp}
\end{center}
\end{figure}

The determination of the shower geometry is commonly performed in
two steps in the fluorescence technique~\cite{flyseye}.
First, the ``shower-detector-plane'' (SDP) is determined as the
plane spanned by the (signal-weighted) viewing directions of the
triggered camera pixels (Fig.~\ref{fig-sdp}).
Next, the geometry of the shower within this SDP is reconstructed based
on the correlation between arrival time of the signals and
viewing angle of the pixels projected into the SDP.
The measured time-angle correlation is compared to the one expected
for different shower geometries,
and the best-fit geometry is determined.
For the calculation of the expected time-angle correlation, the 
following function is in use 
(following e.g.\ Ref.\ \cite{bunner,flyseye,sokolsky}):

\begin{equation}
t_{i}  =  t_{0}+\frac{R_{p}}{c_{\rm vac}}\tan \left( \frac{\chi _{0}-\chi _{i}}{2} \right)
\label{arrival_old}
\end{equation}

where $t_i$ is the arrival time of the photons at camera pixel $i$
(usually, a signal-weighted average arrival time is taken from the
time sequence observed in a pixel),
$t_0$ is the time at which the shower axis vector passes the
closest point to the telescope at a distance $R_p$,
$c_{\rm vac}$ is the vacuum speed of light, 
$\chi_0$ is the angle of incidence of the shower axis within the SDP,
and $\chi_i$ is the viewing angle of pixel $i$ within the SDP
(see also Fig.~\ref{fig-sdp}). 
Comparing the expected $t_i$$-$$\chi_i$ correlation to the observed one
($i=1...n$ for $n$ triggered pixels),
the best-fit parameters $R_p$, $t_0$ and $\chi_0$ in
Eq.~(\ref{arrival_old}) are found by a $\chi^2$-minimization.
Together with the SDP derived previously, the shower geometry is
then fully determined and can also be expressed in terms of shower
impact point, arrival direction, and ground impact time.

Eq.\ (\ref{arrival_old}) is derived as follows.
Assuming the fluorescence light to be emitted by a point-like
object moving at $c_{\rm vac}$  along the shower axis vector,
the shower propagation time $\tau_{{\rm shower},i}$
from point $S_i$ to the point at reference time $t_0$ on
the shower axis (Fig.~\ref{fig-sdp}) can be expressed as

\begin{equation}
\tau_{{\rm shower},i} = \frac{R_p}{c_{\rm vac}\cdot \tan(\chi_0 - \chi_i)}~.
\label{4_shower}
\end{equation}

Next, assuming the fluorescence photons to propagate
on straight lines with $c_{\rm vac}$,
the light propagation time $\tau_{{\rm light},i}$ from $S_i$ to the
telescope is

\begin{equation}
\tau_{{\rm light},i} = \frac{R_p}{c_{\rm vac}\cdot \sin(\chi_0 - \chi_i)}~.
\label{4_prop}
\end{equation}

With Eqs.\ (\ref{4_shower}) and (\ref{4_prop}),
and assuming an instantaneous emission of the fluorescence photons
at $S_i$, the expected arrival time $t_i$ (relative to the time $t_0$
of closest approach of the shower to the telescope)
of fluorescence photons at a pixel viewing at an angle $\chi_i$ 
becomes

\begin{eqnarray}
t_{i} & = & t_{0} - \tau_{{\rm shower},i} + \tau_{{\rm light},i} \nonumber \\
                          & = & t_{0} + \frac{R_{p}}{c_{\rm vac}}
   \left( \frac{1}{\sin (\chi_0 - \chi_i)}-\frac{1}{\tan (\chi_0 - \chi_i)}\right) \nonumber  \\
                          & = & t_{0}+\frac{R_{p}}{c_{\rm vac}}\tan \left( \frac{\chi _{0}-\chi _{i}}{2} \right)
% \mbox{ .}
\label{4_arrival_old}
\end{eqnarray}

which equals Eq.\ (\ref{arrival_old}).

Thus, the derivation of Eq.\ (\ref{arrival_old}) for calculating the expected
time-angle correlation is based on the following assumptions:

\begin{itemize}
\item the spatial structure and the propagation of the shower disk can
      be approximated by a point-like object moving at $c_{\rm vac}$,
\item the fluorescence light is produced instantaneously,
\item the fluorescence light propagates with $c_{\rm vac}$,
\item the fluorescence light propagates on a straight line.
\end{itemize}

In this article, we investigate the validity of these assumptions.
The impact of the corrections on reconstruction parameters is studied.
The results are relevant both for observations
with fluorescence telescopes alone and for ``hybrid'' observations where
the shower is registered by fluorescence and surface detectors.

The plan of the paper is as follows.
In Section~\ref{sec-effects}, we discuss the various assumptions
and their corrections individually. In Section~\ref{sec-impact},
the impact on shower reconstruction is studied.
Conclusions are given in Section~\ref{sec-concl}.

%%%%%%%%%%%%%%%%%%%%%%%%%%%%%%%%%%%%%%%%%%%%%%%%%%%%%%%%%%%%%%%%%%%%%

\section{Analysis of individual effects}
\label{sec-effects}

We discuss step-by-step the individual effects given by
\begin{itemize}
\item the spatial structure and speed of the shower disk
      (instead of a point-like object moving with $c_{\rm vac}$),
\item the delayed (instead of instantaneous) fluorescence light
      emission,
\item the reduced propagation speed of light (instead of $c_{\rm vac}$),
\item the bending of light (instead of straight-line propagation).
\end{itemize}

%%%%             Spatial structure of shower disk %%%%%%%%%
\subsection{Spatial structure and speed of shower disk}
\label{subsec-spatial}

To check the assumption of the shower propagating as a point-like object
with $c_{\rm vac}$ on a straight line, one may first regard the
fastest particles during the cascading process. Assuming, as a
rough estimate, an elasticity of 50\% per interaction, the energy
of the leading particle in a hadronic air shower is
$E_{\rm lp} \simeq (E_0/A_0) \cdot 0.5^n$
after $n$ interactions for a primary particle of energy $E_0$
and mass $A_0$.
For $n = X_{\rm max} / \lambda \simeq 10$ (the depth of shower
maximum in units of the hadronic interaction length),
the energy of the leading particle is $\sim$$10^{-3}E_0$ for primary
protons and of order $\sim$$10^{-5}E_0$ for primary iron.
Hence, $E_{\rm lp} > 10^{13}$~eV for primary energies 
$E_0 > 10^{18}$~eV
around shower maximum, which is the most relevant portion of 
the shower development for fluorescence light observations.
In this case, the accumulated time delay of the leading particles
with respect to an imaginary
shower front moving with $c_{\rm vac}$ from the first interaction
to $X_{\rm max}$ is $\ll$1~ns. This is negligible compared
to current timing resolutions of giant shower detectors.
Lateral deflections of these particles due to transverse momenta
in interactions or deflection in the Earth's magnetic field are
also sufficiently small (below $\sim$1~m).\footnote{
Time delay and lateral deflection of the leading particles may
become non-negligible in case of
considerably smaller $E_0$ or larger $n$ (the latter being rather
relevant for ground array observations of near-horizontal showers).}
For the case of UHE shower observations by fluorescence telescopes
we conclude that the {\it fastest} shower particles can in reasonable
approximation be assumed to move 
on a straight line along the shower axis with $c_{\rm vac}$.

The main contribution to the fluorescence signal in the shower,
however, is due to lower-energy secondaries, particularly
electrons and positrons between 0.1~MeV and several
100~MeV~\cite{risseheck}.\footnote{Note that for the energy transfer from
$>0.1$~MeV electrons to fluorescence photons, the production of even 
lower-energy (e.g.\ $< 1$~keV) electrons is important (for instance, 
the cross-section for exciting the main molecular bands 
(cf.\ Section~\ref{subsec-prod}) has a sharp peak at about 20~eV
electron energy). However, the additional delay from this
intermediate step is $\ll 1$~ns and, thus, negligible for this
analysis~\cite{arqueros}.}
These have larger {\it lateral} displacements from the shower axis
and larger {\it longitudinal} time delays with respect to the
shower front.

Concerning the lateral width of the fluorescence shower beam,
about 80\% of the total fluorescence signal is produced within
$\sim$75~m around the shower axis~\cite{risseheck}.
The impact of the finite shower width on the fluorescence reconstruction
and how to correct it, was previously studied in 
detail \cite{gora06}.
It was shown in Ref.~\cite{gora06} that choosing too small a photon
collection angle around the shower axis during reconstruction
can lead to a signal loss and underestimation of
the primary energy in nearby showers.

\begin{figure}[t]
\begin{center}
\includegraphics[width=10.0cm]{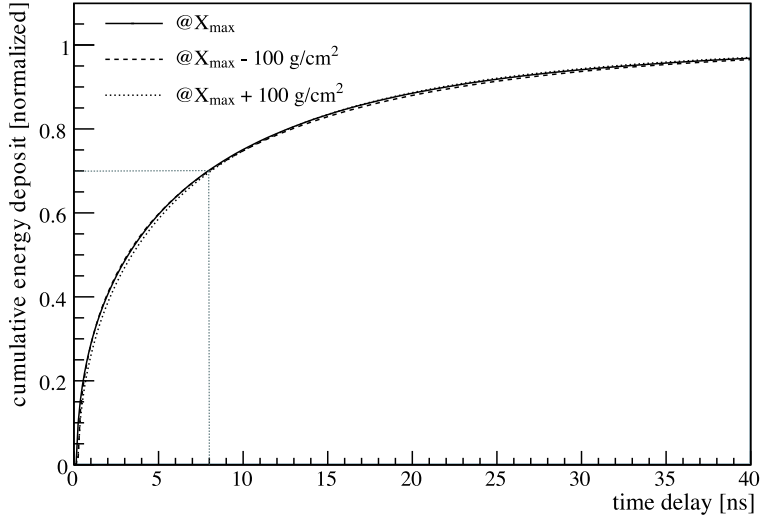}
\caption{
Cumulative energy deposit (normalized to unity) as a
function of time delay with respect to the fastest particle.
The plot refers
to a 10$^{19}$~eV proton at shower maximum (and at
100~g~cm$^{-2}$ smaller/larger depths) and includes particles
within 75~m of the shower axis.
The average time delay is indicated be the dotted line.
Simulations were performed with CORSIKA~\cite{heck} /
QGSJET~01~\cite{qgs01}.
}
\label{fig-cumdeposit}
\end{center}
\end{figure}

Here we study the longitudinal time delay of secondaries using
the CORSIKA code~\cite{heck}.
In Fig.~\ref{fig-cumdeposit} the time delay of secondaries,
weighted according to their contribution to the energy release
into air and thus to the fluorescence signal,
after the arrival time of the first particles is shown
(10$^{19}$~eV shower at maximum, for particles closer than
75~m from the axis; results are practically identical for primary
proton and iron showers). 
One can note a sharp initial increase of the cumulative distribution
(about 50\% of energy is deposited within the first 3$-$4~ns
after the fastest particle),
with a long tail towards larger delays.
The average time delay is $\sim$8~ns, corresponding to a
shower ``thickness'' of a few meters, which is in reasonable
agreement with measurements of particle delays in air showers
(see e.g.\ Ref.~\cite{agnetta}).
As can also be seen in Fig.~\ref{fig-cumdeposit}, the distribution of
time delays changes only marginally with the shower development stage.

The delay of secondaries with respect to the fastest shower particles
results in a small
constant time offset of the observed shower compared to
the assumption of the shower moving with $c_{\rm vac}$.
This might be less relevant for observations with fluorescence
telescopes alone, since in this case, only the relative timing between
the pixels is used to determine the spatial shower geometry.
For hybrid observations, however, usually the arrival time of the
first particle in the ground detector is taken,
while in fluorescence telescopes, usually an average time from a
fit to the signal viewed by a pixel is used.
Then, comparing the timing signals from ground and fluorescence detectors,
the small shift due to the finite shower thickness should be taken into
account.\footnote{For ground detectors located at larger
distances from the shower axis, the curvature of the shower front needs
to be accounted for in addition.}
The precise value of the delay will depend on the specific procedure of 
signal extraction applied during reconstruction. As a rough estimate,
the delay is of order $\tau_{\rm thick} \simeq 5$$-$6~ns.

To summarize, the leading particles in $>$10$^{18}$~eV showers can be
considered to
propagate along the shower axis with $c_{\rm vac}$,
and one can set 
$\tau_{{\rm leading},i} \simeq \tau_{{\rm shower},i}$
with $\tau_{{\rm shower},i}$ given by Eq.~(\ref{4_shower}).
Compared to these particles, the secondaries relevant for the
fluorescence light are slightly delayed due to the finite shower thickness
by $\tau_{\rm thick}$, i.e.\ this term has to be added on the r.h.s.\
of Eq.~(\ref{4_arrival_old}).

%%%%             Fluorescence light production          %%%%%%%%%
\subsection{Fluorescence light production}
\label{subsec-prod}

During propagation, the shower particles excite and ionize
air molecules. Fluorescence light is then emitted by de-excitation
and recombination. Most of the fluorescence light originates from transitions 
from the second positive system (2P) of molecular nitrogen N$_2$ and
the first negative system (1N) of ionized nitrogen molecules~\cite{bunner}.

Typical excitation times are of the order
$\sim$$10^{-6}$~ns~\cite{waldenmaier06} and negligible for current
fluorescence telescopes.
De-excitation times, in turn, can exceed 30~ns.
Depending on the local atmospheric conditions and on the specific
transition system, quenching processes (radiationless transitions by
collisions with other mole\-cu\-les) can substantially reduce the
mean de-excitation time of the radiative processes.

The total reciprocal lifetime 
$1/\tau_{\nu^{\prime}}(p,T)$ of an electronic vibrational state
$\nu^{\prime}$ can be expressed as a function of
pressure $p$ and temperature $T$ as (see e.g.\ Ref.~\cite{waldenmaier07}
and references therein)

\begin{equation}
 \frac{1}{\tau_{\nu^{\prime}}(p,T)} 
 = \frac{1}{\tau_{0_{\nu^{\prime}}}}\left( 1+\frac{p}{p^{\prime}_{\nu^{\prime}}(T)} \right)~.
\end{equation}

Here, $1/\tau_{0_{\nu^{\prime}}}$ is the reciprocal
intrinsic lifetime defined as the sum of all constant
transition probabilities and $p^{\prime}_{\nu^{\prime}}(T)$ is a reference
pressure for a given gas mixture defined as the
pressure where the collisional deactivation constant equals the reciprocal
intrinsic lifetime~\cite{waldenmaier07}.

\begin{figure}[t]
\centering
\includegraphics[width=10cm]{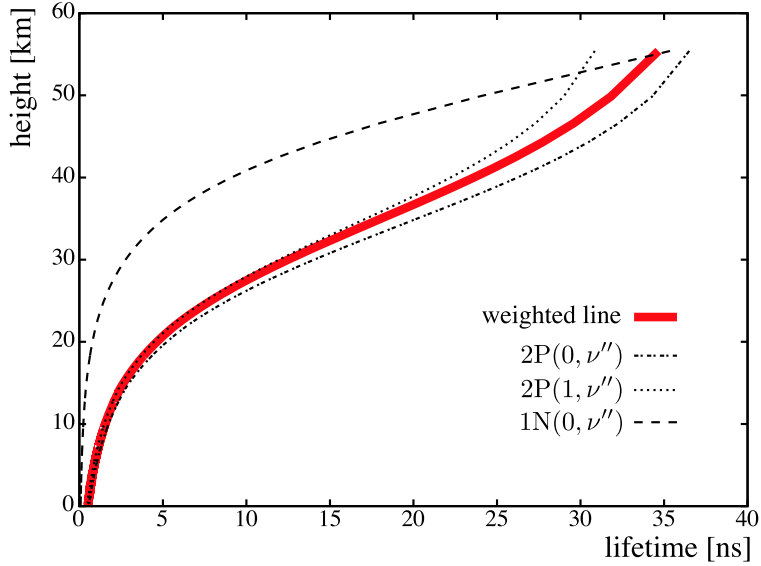}
\caption[Lifetime of the three main sets of bands as a function of height]
{Lifetime of the three main sets of bands as a function of height
a.s.l.\ for dry air.
The thick line shows the averaged lifetime, weighted according to
different intensity fractions.
The width of the line indicates the effect of a change in temperature
by $\pm$40~K.}
\label{life_vs_height}
\end{figure}

Fig.\ \ref{life_vs_height} shows the calculated lifetimes as a function of
height for the three main sets of bands 2P$(0,\nu^{\prime\prime})$,
2P$(1,\nu^{\prime\prime})$ and 1N$(0,\nu^{\prime\prime})$,
% MAR: why \prime\prime ?
assuming dry air ($78.1\%$~N$_2$, $20.9\%$~O$_2$ and $1\%$~Ar) and 
temperature profiles corresponding to the typical conditions at the Auger
Observatory \cite{keilhauer05}. Also shown is the averaged lifetime,
weighting the emission bands according to their relative (altitude dependent)
intensities.
The width of the weighted line indicates the effect of an arbitrary temperature
variation of $\pm$40~K to show the minor dependence of the averaged
lifetime on reasonable variations of the actual atmospheric conditions.
At very high altitudes of 30$-$40~km, the averaged lifetime is 15$-$25~ns. 
With decreasing altitude, the quenching effect reduces the
lifetime; thus, in general, the delay of fluorescence light emission with
respect to the passing shower front is a differential effect that 
changes during the shower development (smaller delay deeper in the
atmosphere).\footnote{Anecdotally, this means the front of fluorescence
light emission can move with an apparent velocity {\it larger} than
$c_{\rm vac}$ through the atmosphere.}
At heights below $\sim$20~km where showers are typically
observed by ground-based observatories, lifetimes of a few ns are
reached.

The average lifetime $\tau_{\rm deexc}$ [in ns]
(weighted line in Fig.\ \ref{life_vs_height})
can in good approximation be parameterized
as a function of height $h$ a.s.l.\ [in m] of the emission point by
\begin{equation}
 \tau_{\rm deexc}(h) = \frac{\tau_{0}}{\alpha\cdot e^{-h/H}+1}~,
\end{equation}
with $\tau_{0}=37.5$~ns, $H=8005$~m and $\alpha=95$. 
The term $\tau_{\rm deexc}(h)$ has to be added to the r.h.s.\ of
Eq.~(\ref{4_arrival_old}).

%%%%         Reduced speed of light                %%%%%%%%%
\subsection{Reduced speed of light}
\label{subsec-photonspeed}

The propagation speed of light $v=c_{\rm vac}/n$ is reduced compared
to the vacuum case by the local index of refraction of air $n$.
The change of $n$ with wavelength is small ($<$3\%)~\cite{bernloehr}
within the fluorescence window of about 300$-$400~nm.
Following Ref.\ \cite{weast63}, the index of refraction can be parametrized
as a function of altitude $h$ as

\begin{equation}
n(h)= 1 + (n_0 -  1)\frac{\rho (h)}{\rho_0}~
\label{6_index_full}
\end{equation}

with the atmospheric density profile $\rho(h)$;
$n_0$ and $\rho_0$ are the reference values at sea level.
The propagation time of refracted light over a small line element $ds$
is then given by $d\tau_{\rm refr} \simeq n(h) ds /c_{\rm vac}$
and for propagation 
between two points at altitudes $h_2$ and $h_1$ ($h_2 > h_1$) by

\begin{equation}
 \tau_{\rm refr} =  \frac{1}{c_{\rm vac} \cos \vartheta}\int_{h_1}^{h_2}n(h)~dh
\end{equation}

with $\vartheta$ being the zenith angle of the propagation 
direction of light.

\begin{figure}[t]
\centering
\includegraphics[width=9cm]{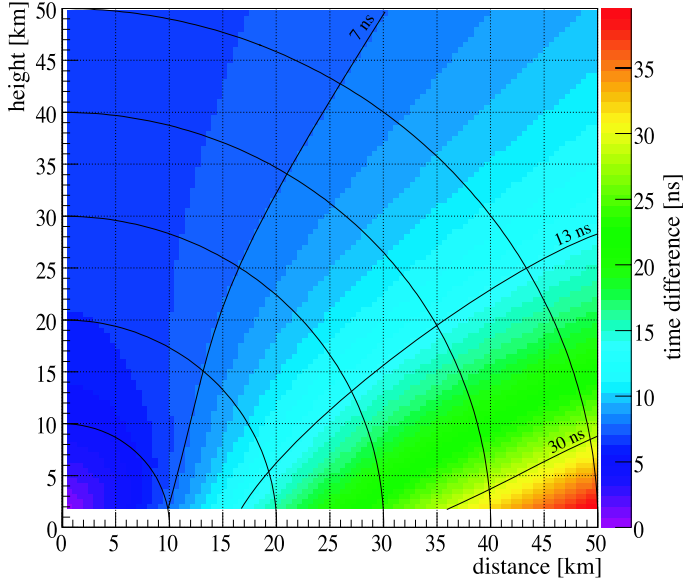}
\caption[Arrival time difference for fluorescence light]
{Arrival time difference $(t_{\rm real}-t_{\rm vacuum})$ due to
the effect of reduced speed of light.
The telescope is placed at 1.4~km a.s.l.\ corresponding
to the altitude of the Auger telescope station ``Los Leones.''
}
\label{diff_time_speed}
\end{figure}

In Fig.~\ref{diff_time_speed}, the difference of light arrival times
(between the cases of vacuum and reduced speed of light) is shown as
a function of the location of emission point with respect to a telescope.
The parametrization of $\rho(h)$ is taken from Ref.~\cite{keilhauer05}
for the example of the southern Auger Observatory.
As expected, for fixed distance between emission point and telescope,
time differences
grow for propagation closer to ground due to the larger value of $n$.
Differences of 20$-$25~ns or more can occur.
For a single air shower, the effect changes along the longitudinal
shower path, depending also on the relative orientation of shower
axis and telescope. For instance, the time difference along the shower
path typically varies less for showers pointing towards the telescope.

In Eq.~(\ref{4_arrival_old}), $\tau_{{\rm light},i}$ is replaced by
$\tau_{{\rm refr},i}$.
For convenience, one can express $\tau_{{\rm refr},i}$ using
Eq.~(\ref{4_prop}) by replacing $c_{\rm vac}$
with $c_{{\rm refr},i} = s / \tau_{\rm refr,i}$,
defined as the effective speed of refracted light along the path of
length $s$ between emission point and telescope.

%

%%%%             Bending of light          %%%%%%%%%
\subsection{Bending of light}
\label{subsec-photonbending}

\begin{figure}[t]
\centering
\includegraphics[width=6.5cm]{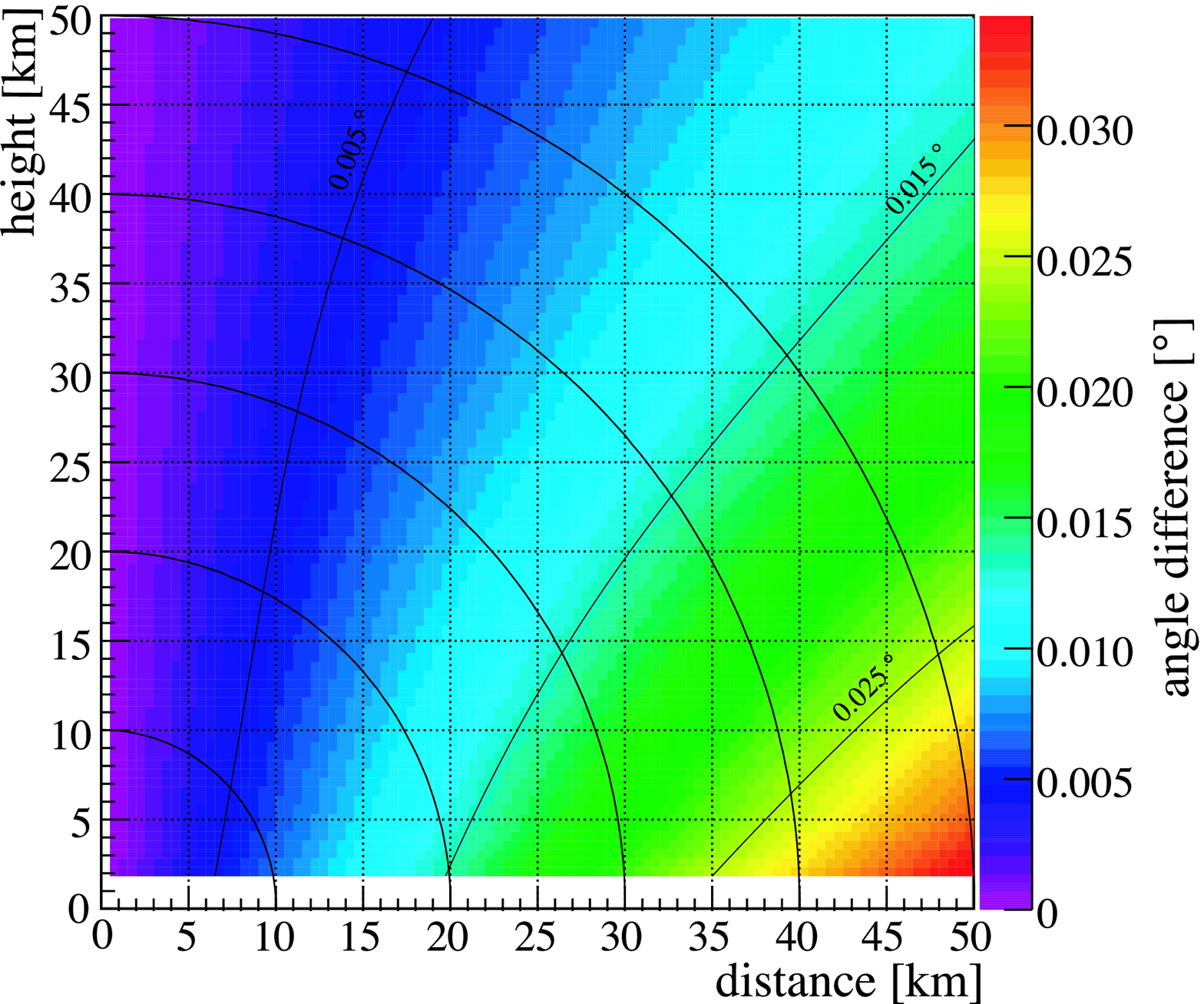}
\includegraphics[width=6.5cm]{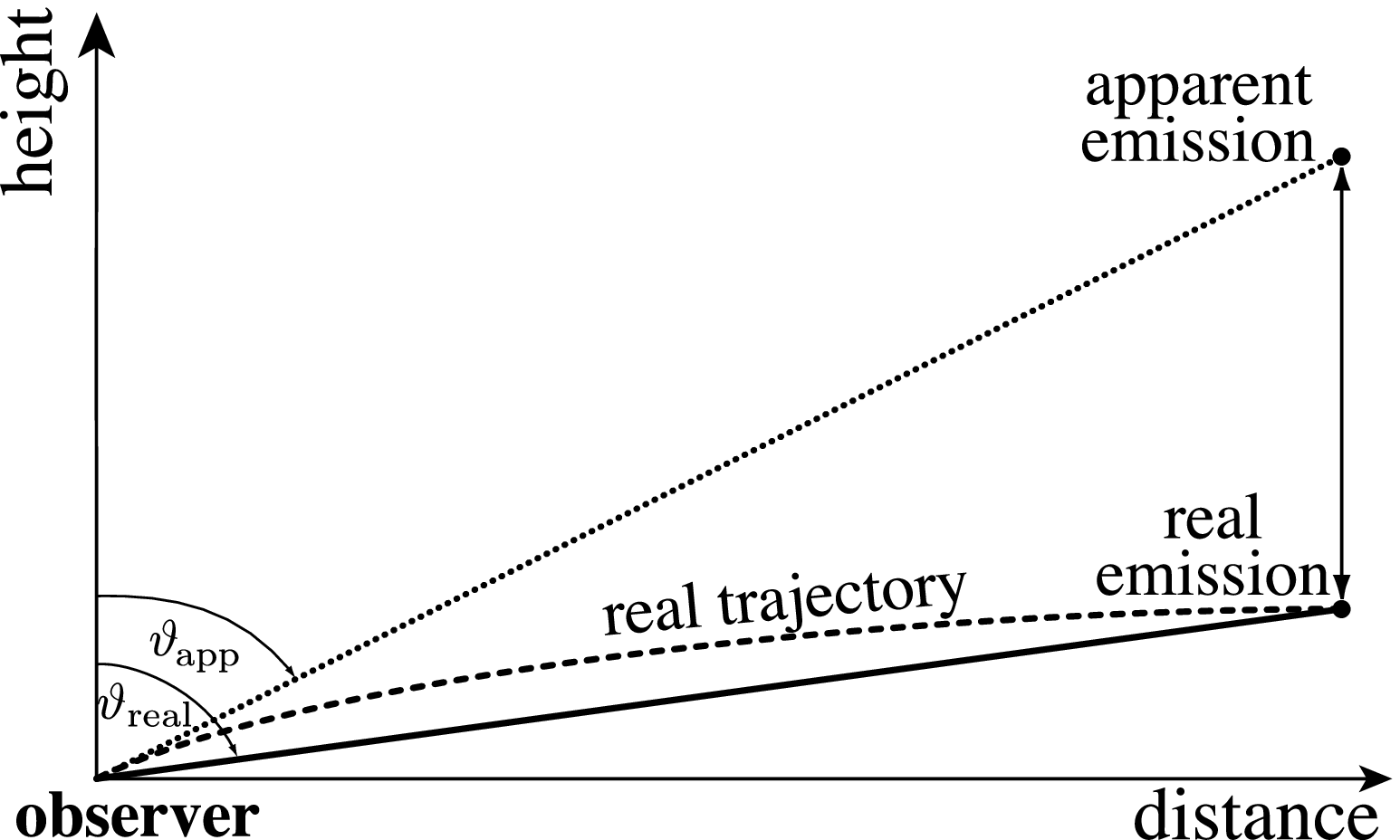}
\caption{Zenith angle difference
$\Delta \vartheta = \vartheta_{\rm real} - \vartheta_{\rm app}$ 
between direct and curved path due to light refraction as a function
of the location of the emission point relative to the telescope.
The telescope is placed at 1.4~km a.s.l.\ corresponding
to the altitude of the Auger telescope station ``Los Leones.''
}
\label{angle_diff}
\end{figure}

Due to refraction, the emitted light propagates on a bent trajectory.
In turn, the direction of the incidence angle of the observed light does
not point towards the real emission point, see Fig.~\ref{angle_diff}
(right).
More specifically, the zenith angle of down-going light is continuously
reduced during propagation.\footnote{We consider here only the case of 
a stable atmosphere with a standard decrease of $\rho(h)$ and $n(h)$ 
with height as given by Eq.~(\ref{6_index_full}) and Ref.~\cite{keilhauer05}.
We note, however, that the path of refracted light can become more
complicated for specific atmospheric conditions such as atmospheric
inversion, or in case of a strongly radiating ground leading to a local
heating of air. The impact of the latter on the fluorescence technique
might be reduced due to the fact that observations are only performed
well ($\sim$1$-$2~h) after / before sunset; also, the shower path very
close to ground usually is below the field of view of the telescope
($\sim$1$^\circ$ elevation of lower edge of field of view).}
%The angle of incidence $\theta_r$  of refracted light follows from
%$ n_r \sin\theta_r =  n_0 \sin\theta_0 $
%(Snell's law)
%where $n_r$ is the local index of refraction, and $n_0$ and $\theta_0$
%are the initial index of refraction and angle of incidence at emission,
%respectively.
The zenith angle difference 
$\Delta \vartheta = \vartheta_{\rm real} - \vartheta_{\rm app} \ge 0$
between the observed light direction $\vartheta_{\rm app}$
(towards the
apparent emission point) and the straight-line direction $\vartheta_{\rm real}$
(towards the real emission point) has been calculated from ray tracing
with $n(h)$ from Eq.~(\ref{6_index_full}); it is shown in
Fig.~\ref{angle_diff} (left) as a function of the position
of the emission point in the atmosphere relative to the telescope.
%\footnote{As a practical parametrization $\Delta \vartheta$ can be
%approximated by 
%$\Delta \vartheta = (\vartheta_{\rm real} - \arcsin((n_1/n_2)\cdot \sin\vartheta_{\rm real}))/(2-h/a)$, 
%where $n_1$ ($n_2$) is the index of refraction of the emission 
%(observation) point, $h$ the emission height in km a.s.l.\ and
%in the considered case of a telescope at 1.4~km a.s.l.\ $a=50$~km.}
% which is the mismatch in pointing; here done by ray tracing
As an example,
an angular difference of 0.02$^\circ$ implies a $\sim$12~m upward shift
of the apparent emission point for a vertical shower at 30~km distance
which corresponds to a $\sim$40~ns shift in time.
These shifts change over the longitudinal viewing direction towards
an air shower.
In case of hybrid observations where timing signals of fluorescence
and ground detectors are combined,
the impact time on ground estimated from the telescopes
will be delayed compared to the actual one.

For a vertical shower, or, more generally, for showers with
$\vartheta_{\rm SDP} = 90^\circ$ (cf.\ Fig.~\ref{fig-sdp}),
$\chi_i$ in Eq.~(\ref{4_arrival_old}) is just reduced by
$\Delta \vartheta$, as the refracted light direction still points
towards the actual shower axis.
In general, however, this effect slightly shifts the
refracted light signals out of the actual SDP, and this shift
usually changes along the shower path. 
Thus, the apparent SDP (which, in fact need not be a ``plane'' anymore)
may slightly be tilted compared to the real one.
To still permit the practical approach of fitting the best shower geometry
within a plane only (instead of testing the full phase space),
the projected shift $\Delta \vartheta \cdot \sin \vartheta_{\rm SDP}$
is taken as a correction. 
Thus, in Eq.~(\ref{4_arrival_old}), $\chi_i$ is replaced by
$\chi _{{\rm refr},i} \simeq \chi_i - \Delta \vartheta_i \cdot \sin \vartheta_{\rm SDP}$
where $\chi _{{\rm refr},i}$ denotes the effective viewing direction
of pixel $i$ due to refraction.
To account for the possible slight tilt of the apparent SDP, 
which is expected to be no larger than 
$\Delta \vartheta^{\rm max} \simeq $ (few times) 0.01$^\circ$,
the best-fit SDP might be found in an iterative procedure.

Finally, we note that the additional time delay due to the increased,
bent path length compared to the straight-line connection
(see sketch in Fig.~\ref{angle_diff}) is $\ll$1~ns and can thus
be neglected.

%%%%%%%%%%%%%%%%%%%%%%%%%%%%%%%%%%%%%%%%%%%%%%%%%%%%%%%%%%%%%%%%%%%%%

\section{Impact on shower reconstruction}
\label{sec-impact}

Taking the discussed effects into account, Eq.~(\ref{arrival_old})
is finally replaced by

\begin{eqnarray}
t_{i}  =  t_{0}
   - \frac{R_{p}}{c_{\rm vac}} \frac{1}{\tan (\chi _{0}-\chi _{{\rm refr},i})} 
   + \frac{R_{p}}{c_{{\rm refr},i}} \frac{1}{\sin (\chi _{0}-\chi _{{\rm refr},i})} 
%\nonumber \\
   + \tau_{\rm thick} 
   + \tau_{{\rm deexc},i} 
% oder:
%   + \Delta\tau_{\rm deposition}
%   + \Delta\tau_{\rm emission}
\label{arrival_new}
\end{eqnarray}

The index $i$ indicates that these quantities, for a given shower geometry,
depend on the viewing direction of pixel $i$.
One caveat, as discussed in Sec.~\ref{subsec-photonbending}, is that the
bending of light slightly changes the apparent SDP (within which the
angles $\chi _{0}$ and $\chi _{{\rm refr},i}$ are defined).
It is worthwhile to note that all correction terms depend only on shower
geometry but not on shower physics such as the primary particle type,
which facilitates their application in shower reconstruction codes.
$\tau_{\rm thick}$ can, to a good degree, be treated as a constant; 
$\tau_{{\rm deexc},i}$ depends on the altitude of the emission point;
and ${c_{{\rm refr},i}}$ and $\chi _{{\rm refr},i}$ depend on the
locations of emission point and telescope.

\begin{figure}[t]
\begin{center}
\includegraphics[width=5.5cm]{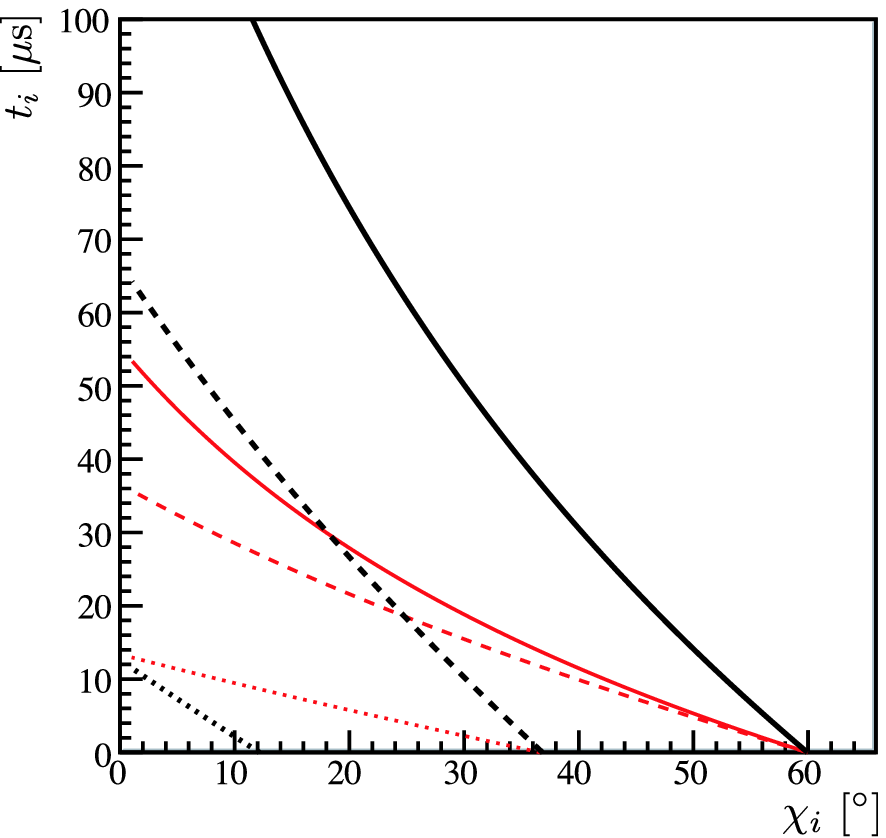}
\includegraphics[width=5.5cm]{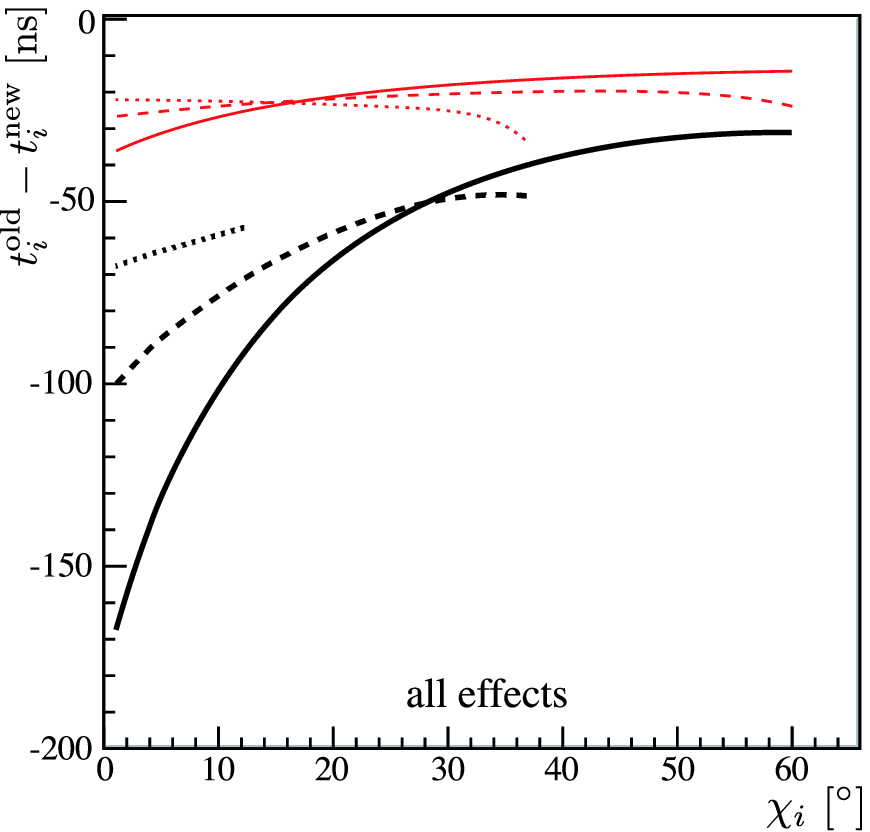}
\includegraphics[width=5.5cm]{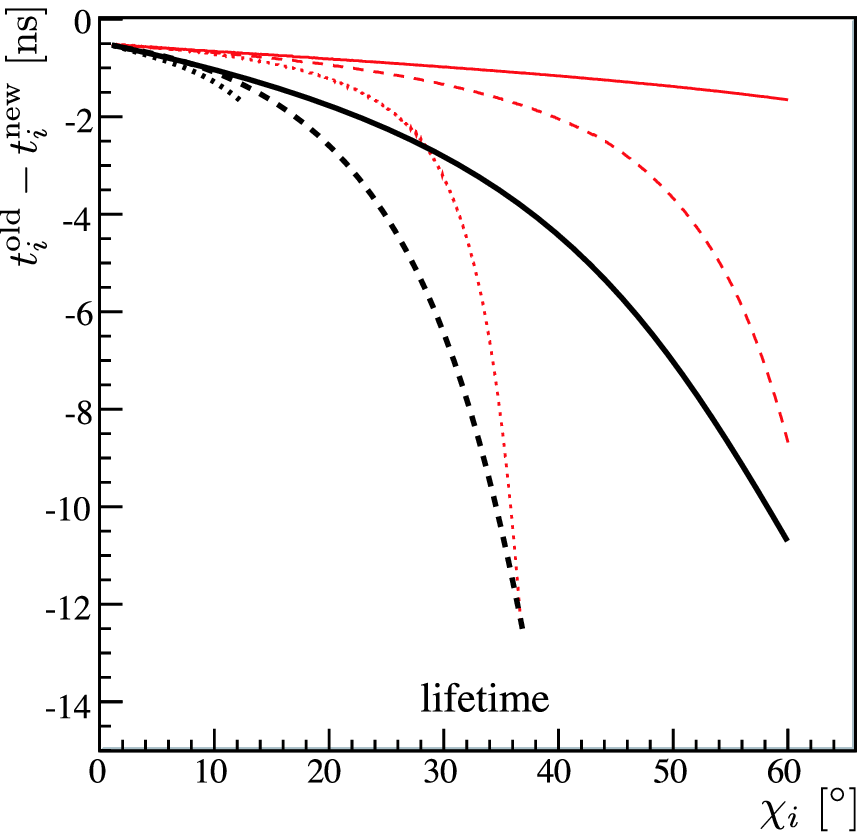}
\includegraphics[width=5.5cm]{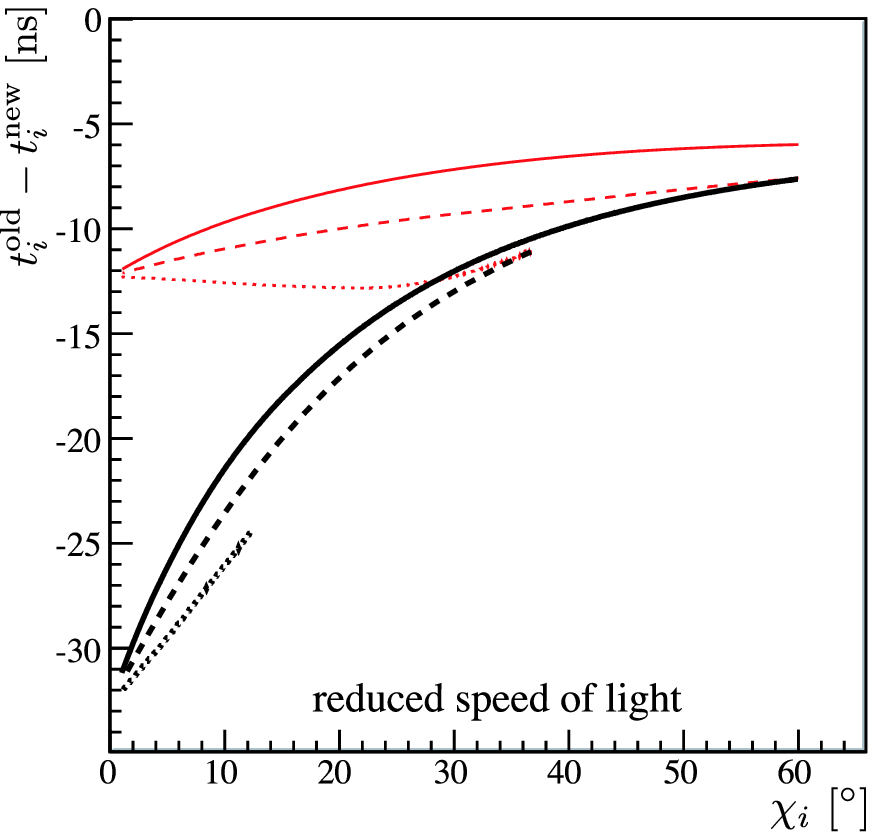}
\includegraphics[width=5.3cm]{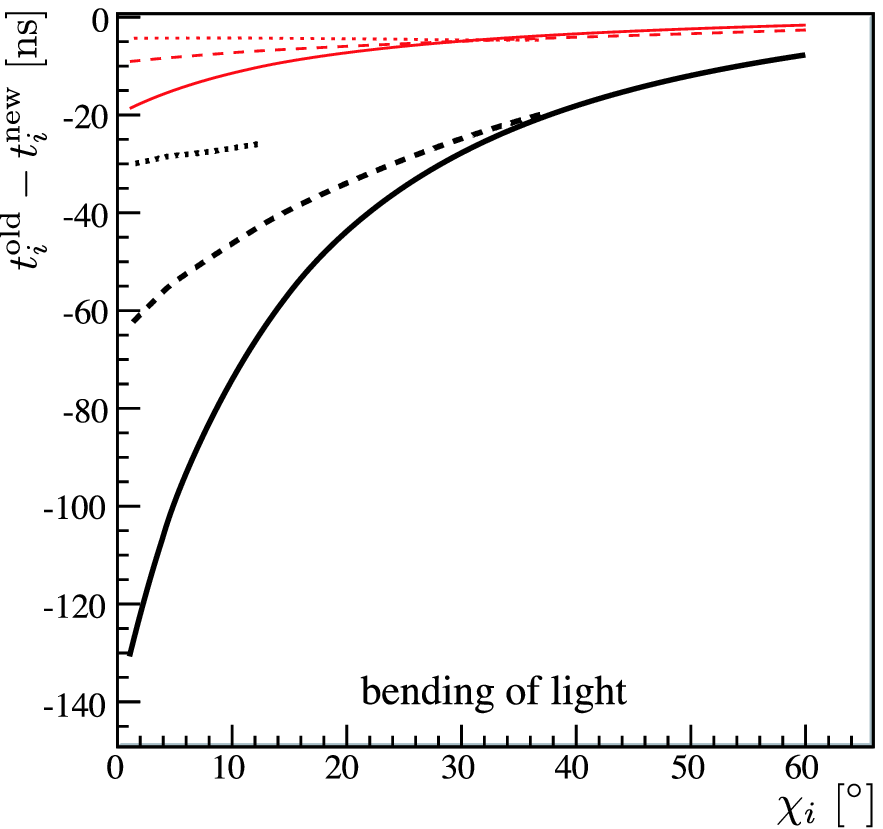}
\includegraphics[width=5.7cm]{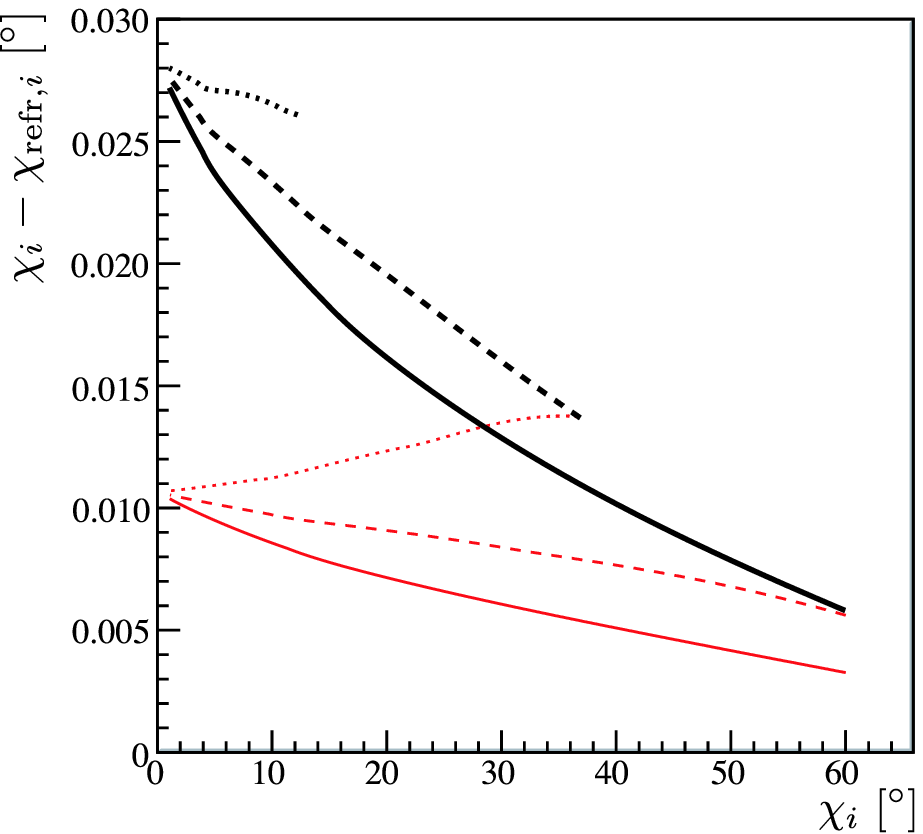}
\caption{Upper left plot: light arrival time $t_i$ vs.\ light arrival
angle (or pixel viewing direction) $\chi_i$ for different
shower geometries (thick black (thin red) lines: shower impact point at
40~km (15~km) distance from the telescope; shower inclination
$\chi_0$ = 130$^\circ$ (solid), 90$^\circ$ (dashed), 50$^\circ$ (dotted);
in all cases $\vartheta_{\rm SDP} = 90^\circ$; shower track shown
up to 50~km distance from the telescope).
Upper right to lower left plot: differences in expected
light arrival time for the given shower geometries
between old and new reconstruction including all effects
(upper right) and for individual effects as assigned.
Lower right plot: differences of viewing angles towards apparent and actual
emission point due to refraction.
}
\label{fig-tchiall}
\end{center}
\end{figure}

The time shifts introduced by the various effects along the
viewing direction $\chi_i$ towards the shower are displayed
in Fig.~\ref{fig-tchiall} for different shower geometries.
The distance between impact point and telescope were fixed to
15~km (thin line) and 40~km (thick line), and for each distance
three different shower inclinations of
$\chi_0 = 50^\circ, 90^\circ, 130^\circ$ are considered. 
Here, for simplicity $\vartheta_{\rm SDP} = 90^\circ$
is taken such that $|90^\circ - \chi_0|$ is identical to the shower
zenith angle. 
In this case, the effect from light bending is minimized concerning
the change of the SDP and maximized concerning 
$ \chi_i - \chi _{{\rm refr},i}$.

In Fig.~\ref{fig-tchiall} (a), the overall shapes of
$t_i$ vs.\ $\chi_i$ are given, which differ for the 
different geometries.
The shift of the arrival times, compared to the previous approach,
is shown in Fig.~\ref{fig-tchiall} (b)
when taking all effects into account. 
The contributions from the individual effects are provided
in Figs.~\ref{fig-tchiall} (c)$-$(e). For the bending of light,
in Fig.~\ref{fig-tchiall} (f), also the shift between apparent
and effective viewing angle is given.
One sees that the time delays are geometry dependent and can reach,
and even exceed, 50$-$100~ns.

One also sees in Fig.~\ref{fig-tchiall} that the time delays change along the shower
track in an individual event.
When reconstructing the shower as a whole, the fitting procedure
then minimizes the overall $\chi^2$ by adjusting simultaneously
$R_p$, $t_0$ and $\chi_0$. To investigate the effective impact of
the corrections on the final reconstruction parameters,
events were generated using CORSIKA~\cite{heck} with the
hadronic interaction model QGSJET~01~\cite{qgs01}.
The shower sample consists of proton induced showers with
energies of $10^{18}$, $10^{19}$ and $10^{20}$~eV
and zenith angles of 0, 45 and 60~deg (100 events per combination
with random azimuth angles).
The detector simulation and the event reconstruction was performed
using the Auger software package described in~\cite{offline,offline2}.
These data were reconstructed with and without accounting
for the discussed effects; or, more specifically, using
once Eq.~(\ref{arrival_old}) and once Eq.~(\ref{arrival_new})
in the reconstruction, and comparing the differences.

\begin{figure}[t]
\centering
\includegraphics[width=4.2cm]{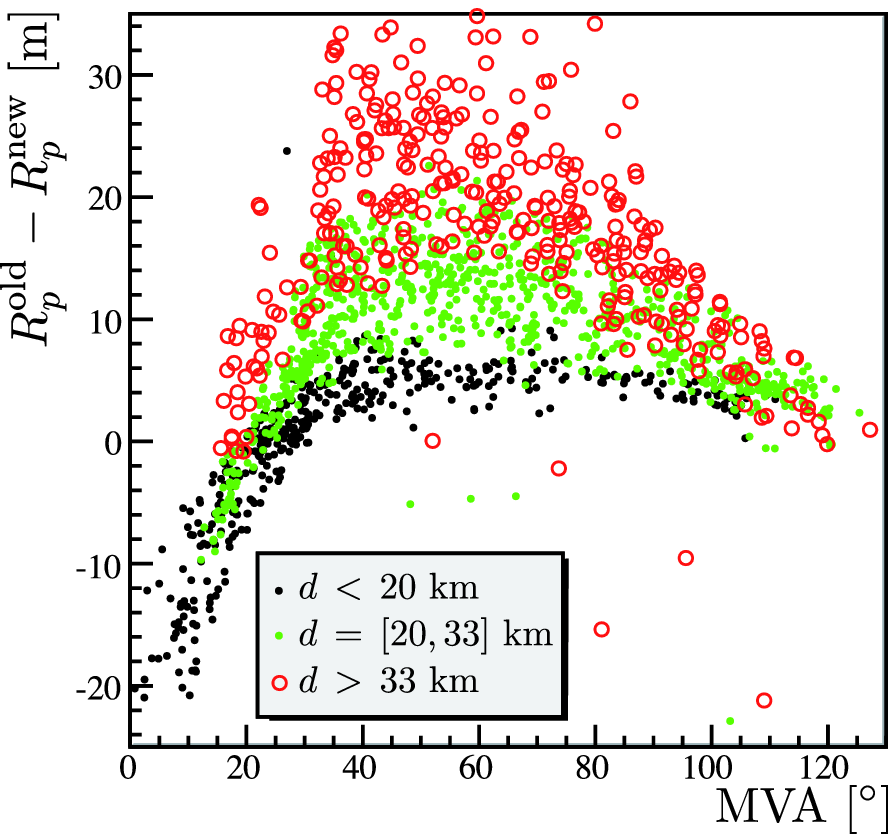}
\includegraphics[width=4.2cm]{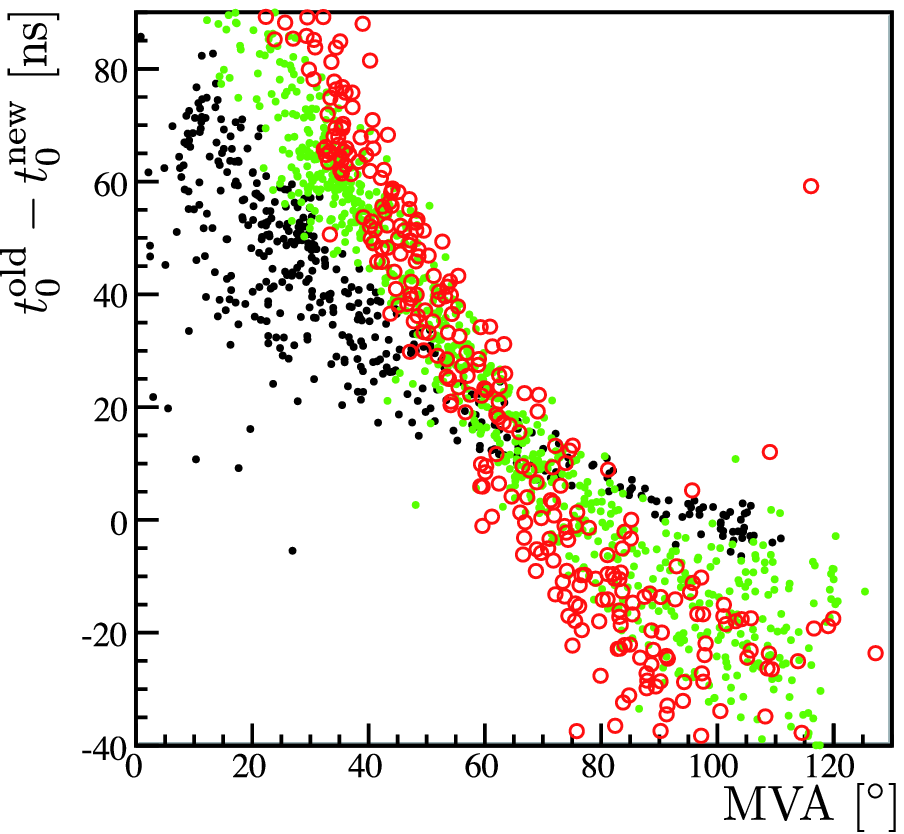}
\includegraphics[width=4.2cm]{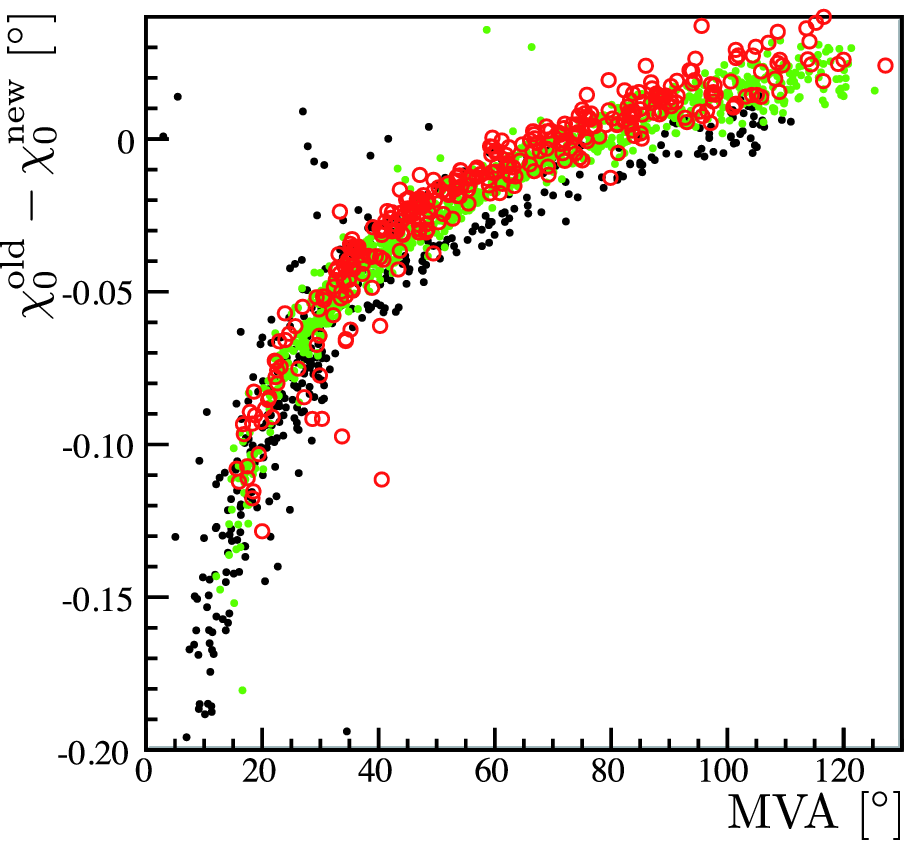}
\caption{Scatter plot of differences between old and new reconstruction
in the geometry parameters $R_p$, $t_0$ and $\chi_0$
as a function of the minimum viewing angle. The events are divided
according to their distance $d$ of the impact point to the telescope
(black dot: $d<20$~km, green dot: $d=[20,33]$~km and red open circle: $d>33$~km).}
\label{sim_data}
\end{figure}
\begin{figure}[t]
\centering
\includegraphics[width=7cm]{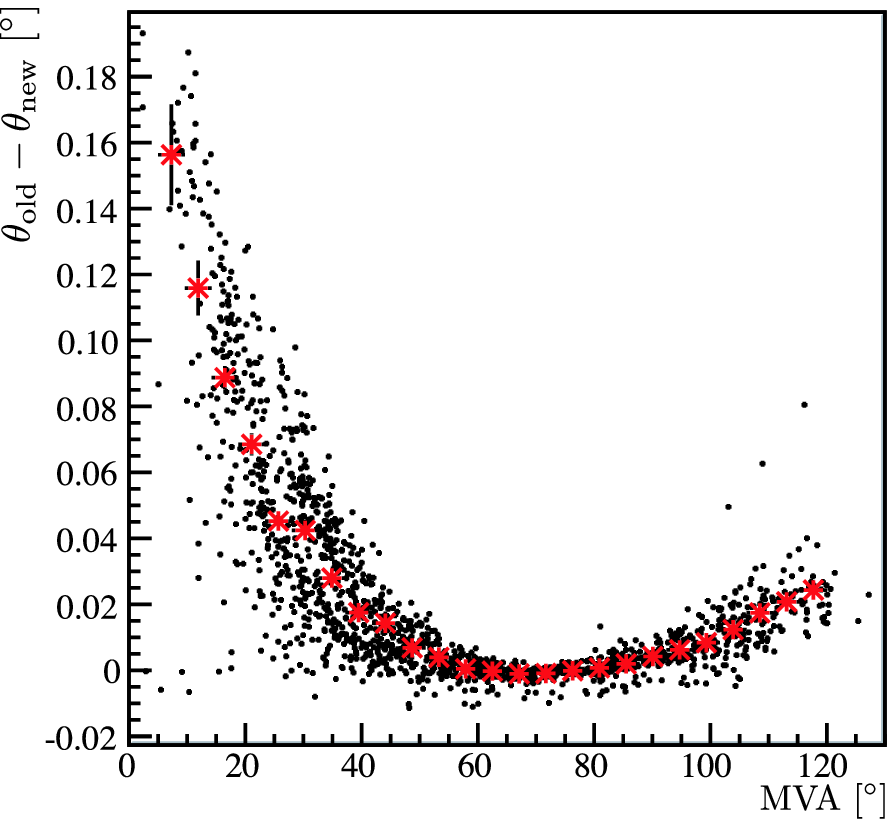}
\caption{Differences between old and new reconstruction in shower
zenith angle as a function of the minimum viewing angle
(dots indicate individual events, red stars the average value).}
\label{sim_dtheta}
\end{figure}

As the time delays from individual parts of the shower were found to
depend on distance and relative orientation of the shower to the
telescope, we plot in Fig.~\ref{sim_data} the change in the parameters $R_p$,
$t_0$ and $\chi_0$ as a function of the minimum viewing angle (MVA),
divided in different distance bins.
The MVA is defined as the smallest angle under which the reconstructed
air shower is seen by the telescope.
Some dependence on MVA and distance can indeed be seen, as expected
also from projection effects
(for instance, a given angular offset in viewing direction
leads to a larger shift along the shower axis for showers pointing
towards the telescope than for vertical showers),
or from an accumulation of certain effects with distance
(such as the time delay due to the reduced speed of light).
The actual impact of the corrections on an individual event is more
complex, however, and has some
dependence also on parameters other than geometry. For instance,
the shower track of a higher-energy event can be observed out to
larger altitudes due to the increased light output, such that these
parts of the shower track can also contribute to the geometry fit.
Thus, an {\it a posteriori} correction of the geometry parameters
determined with Eq.~(\ref{arrival_old}) is not straightforward and would neglect
individual event properties.

% Note that in the simulation chain the propagation speed
% of the fluorescence light is (until now) $c_{\rm vac}$ and hence
% the reconstruction with (vacuum) $c_{\rm vac}$ is consistent
% in contrast to real data. One can see that there is a strong
% dependence for low MVA (shower pointing towards the telescope).
For values of the MVA larger than 40$-$50$^\circ$, deviations up to about 
15~m (in  $R_p$), 30~ns (in $t_0$), and 0.04$^\circ$ (in $\chi_0$) are observed.
Differences in $R_p$ are typically larger in case of more distant showers.
At smaller values of the MVA (showers pointing towards the telescope),
also larger deviations are possible,
e.g.\ shifts in $\chi_0$ of 0.1$^\circ$ or more.
In terms of differences in arrival directions (the relevant
quantity for directional source searches), differences are typically
around 0.05$^\circ$, but can exceed 0.1$^\circ$. A systematic
shift can be noted to slightly overestimate the shower zenith angles
when neglecting the discussed effects, see Fig.~\ref{sim_dtheta}.
Shifts in energy are usually small ($\simeq$ 0.5$-$1\% on average).
Reconstructed values for the depth of shower maximum are 
typically changed by 2$-$3~g~cm$^{-2}$, with a trend of the
corrected $X_{\rm max}$ values being increased, and with
larger corrections (5~g~cm$^{-2}$ and more) towards smaller
values of MVA.

%%%%%%%%%%%%%%%%%%%%%%%%%%%%%%%%%%%%%%%%%%%%%%%%%%%%%%%%%%%%%%%%%%%%%

\section{Conclusion}
\label{sec-concl}

The assumptions used in the ``classical'' function of
Eq.~(\ref{arrival_old}) for reconstructing
the shower geometry from fluorescence light observations 
were investigated.
The finite shower thickness leads to an energy deposition in air by
secondaries which is delayed, compared to the shower front, by about
5$-$6~ns (with some dependence on the specific light collection
algorithm employed).
The emission of fluorescence light is further delayed due to the finite
lifetime of the transitions which, due to quenching, is altitude
dependent. Typical values are a few nanoseconds up to 20~km height, and
$>$15~ns for heights above 30~km. 
The propagation speed of light is reduced by the index of refraction
of air. The delay, compared to a propagation with vacuum speed of
light, depends on the locations of emission point and telescope,
and can exceed 20$-$25~ns.
Finally, another effect of refraction is the bending of light,
which also depends on the locations of emission point and telescope.
Angular differences between the apparent and actual emission point 
of 0.02$^{\circ}$ can occur, which may correspond to time shifts
of several 10~ns.
This effect can also lead to a slight tilt of the SDP.

All these corrections can be considered as geometrical ones, i.e.\
they are independent of specific properties of the individual showers other
than their geometry.
The corrected function for geometry reconstruction is given in
Eq.~(\ref{arrival_new}).
Compared to the previous approach, which assumed maximum propagation
speed of both light and particles as well as no other delays, the
effects of delayed timing (including the effect of bending of light)
accumulate.
In total, differences of up to $\sim$100~ns in
estimated light arrival time are possible.
Air shower experiments with comparable, or better, time resolution
should take these effects into account.
This refers both to data reconstruction and to implementing these
effects in the shower$-$detector simulation.
In terms of overall shower reconstruction parameters,
corrections are typically 0.03$-$0.05$^\circ$ in arrival direction
(with a systematic trend of overestimating the zenith angle
when neglecting the effect),
$\simeq$ 0.5$-$1\% in energy and 2$-$3~g~cm$^{-2}$ in $X_{\rm max}$,
but may in some cases exceed 0.1$^\circ$ and 5~g~cm$^{-2}$.
This is to be compared to typical reconstruction accuracies of 
$\sim$0.6$^\circ$ (directional resolution) \cite{bonifazi05} and
$\sim$11~g~cm$^{-2}$ (systematic $X_{\rm max}$ uncertainty) \cite{unger07}
in case of Auger hybrid events.

The increase in computing time for event reconstruction is modest,
particularly when applying the corresponding corrections only when
approaching convergence in the minimization process (increase of
$\sim$20\% or less, depending on implementation).
Some of the effects investigated in this work might be relevant also for
shower detection techniques other than fluorescence telescope observations
at ultra-high energy, e.g.\ Cherenkov light observations of air showers.

%%%%%%%%%%%%%%%%%%%%%%%%%%%%%%%%%%%%%%%%%%%%%%%%%%%%%%%%%%%%%%%%%%%%%

{\it Acknowledgements:}

We would like to thank our Colleagues from the Pierre Auger
Collaboration for many fruitful discussions, in particular
Fernando Arqueros, Jose Bellido, Bruce Dawson, Philip Wahrlich 
and the members
of the Auger group at the University of Wuppertal.
Figs.\ \ref{sim_data} and \ref{sim_dtheta} were produced using 
Auger software packages~\cite{offline,offline2}.
This work was partially supported by the German Ministry for
Research and Education (Grant 05 CU5PX1/6).

\end{document}